# Convex Cauchy–Schwarz Independent Component Analysis for Blind Source Separation

Zaid Albataineh and Fathi M. Salem

*Abstract*—We present a new high-performance Convex Cauchy–Schwarz Divergence (CCS-DIV) measure for Independent Component Analysis (ICA) and Blind Source Separation (BSS). The CCS-DIV measure is developed by integrating convex functions into the Cauchy–Schwarz inequality. By including a convexity quality parameter, the measure has a broad control range of its convexity curvature. With this measure, a new CCS–ICA algorithm is structured and a non-parametric form is developed incorporating the Parzen window-based distribution. Furthermore, pairwise iterative schemes are employed to tackle the high dimensional problem in BSS. We present two schemes of pairwise non-parametric ICA algorithms, one is based on gradient decent and the second on the Jacobi Iterative method. Several case-study scenarios are carried out on noise-free and noisy mixtures of speech and music signals. Finally, the superiority of the proposed CCS–ICA algorithm is demonstrated in "case-study" metric-comparison performance with FastICA, RobustICA, convex ICA (C-ICA), and other leading existing algorithms.

## I. INTRODUCTION

Blind Signal Processing (BSP) is one of the most challenging and emerging areas in signal processing. BSP remains a very important area of research and development in many domains, e.g. biomedical engineering, image processing, communication system, speech enhancement, remote sensing, etc. BSP techniques do not assume full *apriori* knowledge about the mixing environment, source signals, etc. BSP includes three major areas: Blind Signal Separation (BSS), Independent Component Analysis (ICA), and Multichannel Blind Deconvolution (MBD) [2], [3].

In the following, we provide a focused and a brief overview. ICA is considered a key factor of BSS and unsupervised learning algorithms [1], [2]. ICA specializes to Principal Component Analysis (PCA) and Factor Analysis (FA) in multivariate analysis and data mining, corresponding to second order methods, in which the components are in the form of Gaussian distributions [6 - 9], [1], [2]. However, ICA is a technique that includes higher order statistics (HOS) where the goal is to represent a set of random variables as a linear transformation of statistically independent components.

ICA techniques are based on the assumption of non-Gaussianity and independence of the sources. Let an $M$ observation vector $x = [x_1, x_2, \ldots x_M]^T$ be obtained from $M$ statistically independent sources $s = [s_1, s_2, \ldots s_M]^T$ by $x = As$, where $A$ is an $M \times M$ unknown invertible mixing matrix. The estimated sources can be modeled by $y = Wx$ where $W$ is a demixing matrix. The goal in ICA is to determine a demixing matrix $W$ to estimate the source signals. ICA uses the non-Gaussianity of sources and a dependency measure to find a demixing matrix $W$. A measure, e.g., could be based on the mutual information [12 - 17], Higher Order Statistic (HOS), such as the kurtosis [6-9], or Joint Approximate Diagonalization [10 - 11]. In other words, the demixed matrix is obtained by optimizing such a contrast function.

Furthermore, the metrics of cumulants, likelihood function, negentropy, kurtosis, and mutual information have been developed to obtain a demixing matrix in different adaptations of ICA-based algorithms [1]. FastICA [8] was developed to maximize non-Gaussianity with relative speed and simplicity. Recently, Comon [7] proposed the Robust Independent Component Analysis (R-ICA) method with better performance. He used a truncated polynomial expansion rather than the output marginal probability density functions to simplify the estimation process. Moreover, in [42], the authors developed the rapid ICA algorithm which takes advantage of multi-step past information with respect to the fixed-point method in order to increase the non-guassianity among the estimated signals. In [10–12], the authors have presented ICA using mutual information. They constructed a formulation by minimizing the difference between the joint entropy and the marginal entropy of different sources.

The so-called convex ICA [13] is established by incorporating a convex function into a Jenson's inequality-based divergence measure. Xu et al [14] used the approximation of the Kullback–Leibler (KL) divergence based on the Cauchy–Schwartz inequality. Boscolo et al. [15] established nonparametric ICA by minimizing the mutual information contrast function and by using the Parzen window distribution.

A new contrast function based on nonparametric distribution was developed by Chien and Chen [16], [17] to construct the ICA algorithm. They used the cumulative distribution function (CDF) to obtain a uniform distribution from the observation data. Moreover, Matsuyama et al. [18]

proposed the alpha divergence approach. Also, the f-divergence was proposed by Csiszár et al. [3], [19]. Alternate studies have presented the nonnegative matrix factorization (NMF) to solve the BSS problem [3], [18], [19]. They took advantage of imposing the nonnegative constraints to minimize and measure the approximation errors. The Euclidean distance (ED-DIV) and KL divergence (Kl-DIV) were used as the error functions for NMF problems in [19].

In addition, the maximum-likelihood (ML) criterion [21] is another tool for BSS algorithms [21]–[23]. It is used to estimate the demixing matrix by maximizing the likelihood of the observed data. Recently, Fujisawa et al. [24] have proposed a very robust similarity measure to outliers which they call the Gamma divergence. In addition, the Beta divergence was proposed in [25] and investigated in [3]. While these approaches vary in computational load, their performance is still in need of more improvements.

In this work, we develop an effective and improved measure of dependency among the signals, and then we construct its corresponding (parametric and non-parametric) ICA algorithms. A novel family of dependency divergence is developed which we name Convex Cauchy Schwarz Divergence (CCS-DIV) -- due to its use of the Cauchy Schwarz Inequality "divergence." We develop this new measure by conjugating a convex function *into* the Cauchy–Schwarz inequality-based divergence measure. This new contrast function has a wide range of effective curvature since it is controlled by a convexity parameter. The corresponding convex Cauchy–Schwarz divergence ICA (CCS–ICA) employs the Parzen window density approximation to distinguish the non-Gaussian structure of source densities. We also present two effective pairwise ICA algorithms: one is based on the gradient descent and the other is based on the Jacobi optimization. The link between CCS_DIV, ED-DIV, KL-DIV and CS-DIV is also shown. The efficacy of the corresponding ICA algorithms based on the proposed CCS-DIV is verified by means of several ICA experiments. This CCS–ICA has succeeded effectively in solving the BSS of speech and music signals with and without additive (Gaussian) noise, and it has shown a high comparative performance outperforming other existing ICA-based algorithms.

In this paper, we adopt the following notations. A matrix is denoted in bold capital letter, e.g., $\boldsymbol{A}$; $\boldsymbol{A}^T$ is the matrix transpose of $\boldsymbol{A}$ and its Frobenius norm is denoted by $\|\boldsymbol{A}\|$. The identity matrix of size n is denoted by $\boldsymbol{I_n}$. A vector is denoted by a bold small letter, e.g., **a**, and a scalar is denoted by a small letter, e.g., $a$. Also, we use MATLAB's notations to express the algorithm.

The paper is organized as follows. Section II presents a brief description of several existing divergence measures. Also, motivates and proposes the new convex Cauchy–Schwarz divergence measure. Section III generates the corresponding CCS–ICA methods and presents a Pairwise CCS–ICA algorithm. Comparative performance simulation results and final conclusions are given in Section IV and Section V, respectively.

## II. A Brief Description of Previous Divergence Measures

Divergence, or the related (dis)similarity, measures play an important role in the areas of neural computation, pattern recognition, learning, estimation, inference, and optimization [3]. In general, they measure a quasi-distance or directed difference between two probability distributions which can also be expressed for unconstrained arrays and patterns. Divergence measures are commonly used to find a distance between two n-dimensional probability distributions, say $\boldsymbol{p} = (p_1, p_2, \ldots p_n)$ and $\boldsymbol{q} = (q_1, q_2, \ldots q_n)$. Such a divergence measure is a fundamental key factor in measuring the dependency among observed variables and generating the corresponding ICA-based procedures.

A *metric* is the distance between two pdfs if the following conditions hold: $(i)\ D(\boldsymbol{p}\|\boldsymbol{q}) = \sum_{i=1}^{n} d(p_i, q_i) \geq 0$ with equality if and only if $\boldsymbol{p} = \boldsymbol{q}$, $(ii)\ D(\boldsymbol{p}\|\boldsymbol{q}) = D(\boldsymbol{q}\|\boldsymbol{p})$ and $(iii)$ the triangular inequality, i.e., $D(\boldsymbol{p}\|\boldsymbol{q}) \leq D(\boldsymbol{p}\|\boldsymbol{z}) + D(\boldsymbol{z}\|\boldsymbol{q})$, for another distribution $\boldsymbol{z}$. Distances which are not a metric are referred to as divergences [3].

This paper is mostly interested in distance-type divergence measures that are separable, thus, satisfying the condition $D(\boldsymbol{p}\|\boldsymbol{q}) = \sum_{i=1}^{n} d(p_i, q_i) \geq 0$ with equality holds if and only if $\boldsymbol{p} = \boldsymbol{q}$. But they are not necessarily symmetric as in condition (ii) above, nor do necessarily satisfy the triangular inequality as in condition (iii) above.

Usually, the vector $\boldsymbol{p}$ corresponds to the observed data and the vector $\boldsymbol{q}$ is the estimated or expected data that are subject to constraints imposed on the assumed models. For the BSS (ICA and NMF) problems, $\boldsymbol{p}$ corresponds to the observed sample data matrix $\mathbf{X}$ and $\boldsymbol{q}$ corresponds to the estimated sample matrix $\mathbf{Y} = \mathbf{WX}$. Information divergence is a measure between two probability curves. In other words, the distance-type measures under consideration are not necessarily a metric on the space $P$ of all probability distributions [3].

Next, we review the most common divergence measures with one-dimensional probability curves.

### A. Previous Divergence Measures

The KL divergence (KL-DIV) [3, 5] is the relative entropy between the joint distributions of two continuous variables $x_1$ and $x_2$ ($p(x_1, x_2)$) and the product of their marginal distributions ($p(x_1)p(x_2)$). KL-DIV is specifically given by

$$D_{KL}(x_1, x_2) = H\big(p(x_1)\big) + H\big(p(x_2)\big) - H\big(p(x_1, x_2)\big) \quad (1)$$

$$D_{KL}(x_1, x_2) = \iint p(x_1, x_2) \log\left(\frac{p(x_1, x_2)}{p(x_1) \cdot p(x_2)}\right) dx_1 dx_2 \quad (2)$$

Where $H$ represents the entropy operator and $D_{KL}(x_1, x_2) \geq 0$, with equality if and only if $p(x_1) = p(x_2)$. This means that the variables become independent of each other. Xu [14] developed the Euclidean divergence (E-DIV) and the Cauchy–Schwarz divergence (CS-DIV) by joining the terms of joint distributions of two variables and their product of marginal distributions into the Euclidean distance and the Cauchy–Schwarz inequality, respectively. The E-DIV and CS-DIV are given respectively as

$$D_E(x_1, x_2) = \iint (p(x_1, x_2) - p(x_1) \cdot p(x_2))^2 \, dx_1 dx_2 \quad (3)$$

$$D_{CS}(x_1, x_2) = \log \frac{\iint p(x_1, x_2)^2 dx_1 dx_2 \cdot \iint p(x_1)^2 \cdot p(x_2)^2 dx_1 dx_2}{[\iint p(x_1, x_2) p(x_1) p(x_2) dx_1 dx_2]^2} \quad (4)$$

where $D_E(x_1, x_2) \geq 0$ and $D_{CS}(x_1, x_2) \geq 0$ and the equality holds if and only if $p(x_1) = p(x_2)$. At equality, the variables are independent of each other. These divergence measures are reasonable contrast functions to be used in developing the ICA methods. Furthermore, the alpha divergence (α-DIV) was developed by Ali & Silvey who defined a family of convex divergence measures which includes the alpha-divergence as a special case in [32] and was later studied by Amari et. al. [2], [3]. This measure is given specifically by:

$$D_\alpha(x_1, x_2, \alpha) = \iint \left[ \frac{1-\alpha}{2} p(x_1, x_2) + \frac{1+\alpha}{2} p(x_1) \cdot p(x_2) - p(x_1, x_2)^{\frac{1-\alpha}{2}} (p(x_1) \cdot p(x_2))^{\frac{\alpha+1}{2}} \right] dx_1 dx_2 \quad (5)$$

Matsuyama [18] introduced the alpha ICA algorithm by using the α-DIV as a contrast function. In the case $\alpha = -1$, the α-DIV is equivalent to the KL-DIV [3], [5]. Csiszár [19] introduced another divergence measure that is called the f-divergence (f-DIV) and is given by

$$D_f(x_1, x_2) = \iint p(x_1, x_2) f\left(\frac{p(x_1, x_2)}{p(x_1) \cdot p(x_2)}\right) dx_1 dx_2 \quad (6)$$

where $f(.)$ denotes a convex function satisfying $f(t) \geq 0$ for $t \geq 0$, and $f(1) = 0$, $\acute{f}(1) = 0$. In addition, Csiszár shows that the α-DIV is a special case of the f-DIV when using the following convex function

$$f(t) = \frac{4}{1-\alpha^2}\left[\frac{1-\alpha}{2} + \frac{1+\alpha}{2}t - t^{\frac{1+\alpha}{2}}\right] \text{ for } t \geq 0 \quad (7)$$

Furthermore, Zhang [26] developed a general divergence function by integrating the α-DIV and the f-DIV functions into the following form:

$$D_Z(x_1, x_2) = \frac{4}{1-\alpha^2}\left\{\frac{1-\alpha}{2} \iint f(p(x_1, x_2)) \, dx_1 dx_2 + \frac{1+\alpha}{2} \iint f(p(x_1) \cdot p(x_2)) \, dx_1 dx_2 - \int f\left(\frac{1-\alpha}{2} p(x_1, x_2) + \frac{1+\alpha}{2} p(x_1) \cdot p(x_2)\right) dx_1 dx_2\right\} \quad (8)$$

Lin [27] developed a Jensen–Shannon divergence (JS-DIV) by using the Shannon entropy $H[.]$ into the Jensen's inequality; the JS_DIV is given by

$$D_{JS}(x_1, x_2) = H(\lambda p(x_1, x_2) + (1-\lambda) p(x_1) p(x_2)) - \lambda H(p(x_1, x_2)) - (1-\lambda) H(p(x_1) p(x_2)) \quad (9)$$

where $0 \leq \lambda \leq 1$ represents a weighting parameter between the joint distribution and the product of their corresponding marginal distributions. $D_{JS}(x_1, x_2) \geq 0$, and the equality holds if and only if $p(x_1) = p(x_2)$. Recently Chien [13] proposed the convex divergence (C-DIV) by using the Jensen's inequality. The C-DIV is developed by combining the convex function $f(.)$ into the Jensen's inequality. The C-DIV is given by

$$D_C(x_1, x_2, \alpha) = \frac{4}{1-\alpha^2}\left\{\iint \lambda \left[\frac{1-\alpha}{2} + \frac{1+\alpha}{2} p(x_1, x_2) - p(x_1, x_2)^{\frac{1+\alpha}{2}}\right] dx_1 dx_2 + (1-\lambda) \iint \left[\frac{1-\alpha}{2} + \frac{1+\alpha}{2} p(x_1) \cdot p(x_2) - (p(x_1) \cdot p(x_2))^{\frac{1+\alpha}{2}}\right] dx_1 dx_2 - \left[\frac{1-\alpha}{2} + \frac{1+\alpha}{2}(\lambda p(x_1, x_2) + (1-\lambda) p(x_1) p(x_2)) - (\lambda p(x_1, x_2) + (1-\lambda) p(x_1) p(x_2))^{\frac{1+\alpha}{2}}\right]\right\} \quad (10)$$

In the case $\alpha = 1$, the C-DIV is equivalent to the JS-DIV. $D_C(x_1, x_2, \alpha) \geq 0$ and the equality holds if and only if $p(x_1) = p(x_2)$, which means they are independent of each other.

Moreover, in [33], the authors introduced the Beta-divergence, which has a dually flat structure of information geometry and is given by [3]

$$D_B^\beta(x_1, x_2, \beta) = \iint \left( p(x_1, x_2)^\beta \frac{p(x_1, x_2)^\beta - (p(x_1) \cdot p(x_2))^\beta}{\beta} - \frac{p(x_1, x_2)^\beta - (p(x_1) \cdot p(x_2))^\beta}{\beta + 1} \right) dx_1 dx_2 \quad (11)$$

where $\beta$ is a real number and it is not equal $-1 \text{ or } 0$.
Note that the $\beta$-DIV is equivalent to the E-DIV, KL-DIV and Itakura-Saito divergence (IK-DIV) when $\beta = 1$, $\beta \to 0$ and $\beta \to -1$, respectively [3].

### B. The Proposed Divergence Measure

While there exist a wide range of measures, performance especially in audio and speech applications still requires improvements. The quality of an improved measure should provide geometric properties for a contrast function in anticipation of a dynamic (e.g., gradient) search in a parameter space of de-mixing matrices. The motivation here is to introduce a simple measure and incorporate controllable convexity in order to control convergence to the optimal solution. To improve the performance of the divergence measure and speed up the convergence, we have conjugated convex function into (not merely applying it to) the Cauchy–Schwarz inequality. In this context, the paper takes advantage of the convexity's parameter, say alpha, to control the convexity in the divergence function and to speed up the convergence in the corresponding ICA and NMF algorithms. Incorporating the joint distribution ($p(x_1, x_2)$) and the marginal distributions ($p(x_1) p(x_2)$) into the convex function $f(.)$ in (7) and conjugating them to the Cauchy–Schwartz inequality yields

$$\left|\langle f(p(x_1, x_2)), f(p(x_1) p(x_2)) \rangle\right|^2 \leq \langle f(p(x_1, x_2)), f(p(x_1, x_2)) \rangle \cdot \langle f(p(x_1) p(x_2)), f(p(x_1) p(x_2)) \rangle \quad (12)$$

where $\langle \cdot , \cdot \rangle$ is an inner product; $f(.)$ is a convex function, e.g.,

$$f(t) = \frac{4}{1-\alpha^2}\left[\frac{1-\alpha}{2} + \frac{1+\alpha}{2}t - t^{\frac{1+\alpha}{2}}\right] \text{ for } t \geq 0 \quad (13)$$

Now, based on the Cauchy–Schwarz inequality a new

symmetric divergence measure is proposed, namely:
$$D_{CCS}(x_1, x_2, \alpha) = \log \frac{\iint f^2(p(x_1,x_2))dx_1dx_2 \cdot \iint f^2(p(x_1) \cdot p(x_2))dx_1dx_2}{[\iint f(p(x_1,x_2)) \cdot f(p(x_1)p(x_2))\, dx_1 dx_2]^2} \quad (14)$$

where, as usual, $D_{CCS}(x_1, x_2, \alpha) \geq 0$ and equality holds if and only if $p(x_1) = p(x_2)$. This divergence function is then used to develop the corresponding ICA and NMF algorithms. Notably, the joint distribution and the product of the marginal densities in $D_{CCS}(x_1, x_2, \alpha)$ is symmetric. This symmetrical property does not hold for the KL-DIV, α-DIV, and f-DIV. We anticipate that it would be desirable in the geometric structure of the search space. Additionally, the CCS-DIV is tunable by the convexity parameter α. In contrast to the C-DIV [13] and the α-DIV [18], the convexity parameter α range is extendable. However, Based on l'Hopital's rule, one can derive the realization of CCS-DIV for the case of $\alpha = 1$ and $\alpha = -1$ by finding the derivatives, with respect to $\alpha$, of the numerator and denominator for each parts of $D_{CCS}(x_1, x_2, \alpha)$. Thus, the CCS-DIV with $\alpha = 1$ and $\alpha = -1$ are respectively given by (15) and (16).

*C. Link to other Divergences:*

This CCS-DIV distinguishes itself from the previous divergences in the literature by incorporating the convex function into (not merely a function of) the Cauchy Shawarz inequality-- in order to guarantee convexity in the new divergence. This paper thus develops a framework for generating a family of dependency measure based on conjugating the convex function into the Cauchy Shawarz inequality. Such convexity is anticipated (as is evidenced by experiments) to reduce local minimum near the optimal solution and enhance searching a non-linear surface of the contrast function. Also, it provides a flexibility of scalability to high dimensional data. The motivation behind this divergence is to render the CS-DIV to be convex similar to the f-DIV. For this work, we shall focus on one convex function $f(t)$ as in (13), and its corresponding CCS-DIVs in (14), (15) and (16). It can be seen that the CCS-DIV, for the $\alpha = 1$ and $\alpha = -1$ cases, is implicitly based on Shannon entropy (KL divergence) and Renyi's quadratic entropy, respectively. Also, it is to show that the CCS_DIVs for the $\alpha = 1$ and $\alpha = -1$ cases are convex functions in contrast to the CS-DIV. (See Fig. 2 and sub-section E next page.)

*D. Geometrical Interpretation of the Proposed Divergence for $\alpha = 1$ and $\alpha = -1$.*

For simplicity, let's define the following terms:

$$V_J = \iint (p(x_1, x_2))^2 dx_1 dx_2$$

$$V_M = \iint (p(x_1)p(x_2))^2 dx_1 dx_2$$

$$V_c = \iint p(x_1, x_2)p(x_1)p(x_2) dx_1 dx_2$$

$$V_{JJ} = \begin{cases} \iint \left\{ \begin{pmatrix} p(x_1,x_2) \cdot \log(p(x_1,x_2)) \\ -p(x_1,x_2) + 1 \end{pmatrix}^2 \right\} dx_1 dx_2 & \alpha = 1 \\ \iint \left\{ \begin{pmatrix} \log(p(x_1,x_2)) \\ -p(x_1,x_2) + 1 \end{pmatrix}^2 \right\} dx_1 dx_2 & \alpha = -1 \end{cases}$$

$$V_{MM} = \begin{cases} \iint \left\{ \begin{pmatrix} p(x_1) \cdot p(x_2) \cdot \log(p(x_1) \cdot p(x_2)) \\ -p(x_1) \cdot p(x_2) + 1 \end{pmatrix}^2 \right\} dx_1 dx_2 & \alpha = 1 \\ \iint \left\{ \begin{pmatrix} \log(p(x_1) \cdot p(x_2)) \\ -p(x_1) \cdot p(x_2) + 1 \end{pmatrix}^2 \right\} dx_1 dx_2 & \alpha = -1 \end{cases}$$

$$V_{CC} = \begin{cases} \iint \left\{ \begin{pmatrix} p(x_1,x_2) \cdot \log(p(x_1,x_2)) \\ -p(x_1,x_2) + 1 \end{pmatrix} \cdot \\ \begin{pmatrix} p(x_1) \cdot p(x_2) \cdot \log(p(x_1) \cdot p(x_2)) \\ -p(x_1) \cdot p(x_2) + 1 \end{pmatrix} \right\} dx_1 dx_2 & \alpha = 1 \\ \iint \left\{ \begin{pmatrix} \log(p(x_1,x_2)) \\ -p(x_1,x_2) + 1 \end{pmatrix} \cdot \\ \begin{pmatrix} \log(p(x_1) \cdot p(x_2)) \\ -p(x_1) \cdot p(x_2) + 1 \end{pmatrix} \right\} dx_1 dx_2 & \alpha = -1 \end{cases}$$

With these terms, one can express the CCS-DIV and the CS-DIV as

$$D_{CCS} = \log(V_{JJ}) + \log(V_{MM}) - 2\log(V_{CC}) \quad (17)$$

$$D_{CS} = \log(V_J) + \log(V_M) - 2\log(V_C) \quad (18)$$

In Fig. 1, we illustrate the geometrical interpretation of the proposed divergence (CCS-DIV), which is equivalent to Cauchy Schwarz Divergence (CS-DIV). Geometrically, we can show that the angle between the Joint pdfs and Marginal pdfs in the CCS-DIV is given as following:

$$\theta_{CCS} = \mathrm{acos}\left(\frac{V_{CC}}{\sqrt{V_{JJ}V_{MM}}}\right) \equiv \theta_{CS} = \mathrm{acos}\left(\frac{V_C}{\sqrt{V_J V_M}}\right) \quad (19)$$

where $acos$ denotes the cosine inverse. As a matter of fact, the convex function $f$ renders the CS-DIV a Convex contrast function for the $\alpha = 1\ and\ \alpha = -1.$ cases

$$D_{CCS}(x_1, x_2, 1) = \log \frac{\left(\iint\left\{(p(x_1,x_2) \cdot \log(p(x_1,x_2)) - p(x_1,x_2) + 1)^2\right\}dx_1dx_2\right) \cdot \left(\iint\left\{(p(x_1) \cdot p(x_2) \cdot \log(p(x_1) \cdot p(x_2)) - p(x_1) \cdot p(x_2) + 1)^2\right\}dx_1dx_2\right)}{[\iint\{(p(x_1,x_2) \cdot \log(p(x_1,x_2)) - p(x_1,x_2) + 1) \cdot (p(x_1) \cdot p(x_2) \cdot \log(p(x_1) \cdot p(x_2)) - p(x_1) \cdot p(x_2) + 1)\}dx_1dx_2]^2} \quad (15)$$

$$D_{CCS}(x_1, x_2, -1) = \log \frac{\left(\iint\left\{(\log(p(x_1,x_2)) - p(x_1,x_2) + 1)^2\right\}dx_1dx_2\right) \cdot \left(\iint\left\{(\log(p(x_1) \cdot p(x_2)) - p(x_1) \cdot p(x_2) + 1)^2\right\}dx_1dx_2\right)}{[\iint\{(\log(p(x_1,x_2)) - p(x_1,x_2) + 1) \cdot (\log(p(x_1) \cdot p(x_2)) - p(x_1) \cdot p(x_2) + 1)\}dx_1dx_2]^2} \quad (16)$$

Moreover, it provides the proposed measure an advantage over the CS-DIV in terms of speed and accuracy, see fig. 2 (a) and (d).

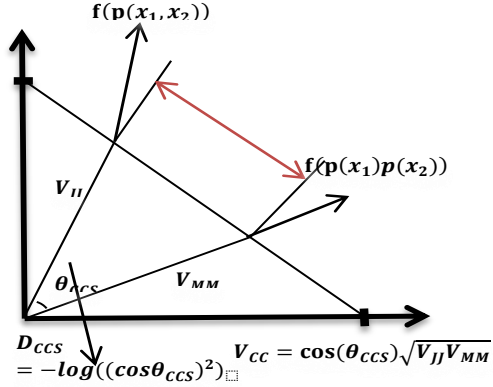

Fig.1: Illustration of Geometrical Interpretation of the proposed Divergence

*E. Evaluation of Divergence Measures*

In this section, the relations among the KL-DIV, E-DIV, CS-DIV, JS-DIV, α-DIV, C-DIV, and the proposed CCS-DIV are discussed. The C-DIV, α-DIV, and the proposed CCS-DIV with $\alpha = 1$ and $\alpha = -1$ are evaluated. Without loss of generality, a simple case is considered. Two binomial variables $\{x_1, x_2\}$ in the presence of the binary events {A, B} are considered as in [13], [14], and [27].

The joint probabilities are $p_{x_1,x_2}(A,A)$, $p_{x_1,x_2}(A,B)$, $p_{x_1,x_2}(B,A)$ and $p_{x_1,x_2}(B,B)$, and the marginal probabilities are $p_{x_1}(A), p_{x_1}(B), p_{x_2}(A)$ and $p_{x_2}(B)$. Different divergence methods are tested by fixing the marginal probabilities of $x_1$, e.g. $p_{x_1}(A) = 0.6$, and $p_{x_1}(B) = 0.4$. And then setting the joint probabilities of $p_{x_1,x_2}(A,A)$ and $p_{x_1,x_2}(B,A)$ free in the intervals (0, 0.6) and (0, 0.4), respectively. Fig. 2 shows the different divergence measures versus the joint probabilities $p_{x_1,x_2}(A,A)$ and $p_{x_1,x_2}(B,A)$. According to fig. 1, all the divergence measures reach the same minimum on the line $p_{x_1,x_2}(A,A) = 1.5 p_{x_1,x_2}(B,A)$, which means that the two random variables become independent. One can observe that the CS-DIV is not a convex function of the pdfs in contrast to CCS-DIV from the graphs in Fig. 2. Furthermore, we then fix the marginal probabilities of $\{x_1, x_2\}$, say at $p_{x_1}(A) = 0.7, p_{x_1}(B) = 0.3, p_{x_2}(A) = 0.5$ and $p_{x_2}(B) = 0.5$, and letting the joint probabilities of $p_{x_1,x_2}(A,A)$ and $p_{x_1,x_2}(B,A)$ be free over the intervals (0, 0.7) and (0, 0.3), respectively. Fig. 3 shows plots of the different divergence measures versus the joint probability $p_{x_1,x_2}(A,A)$. All the divergence measures reach the same minimum at $p_{x_1,x_2}(A,A) = 0.35$, which means that the two random variables become independent. Among these measures, the steepest curve is obtained by the CCS-DIV at $\alpha = -1$. A plausible note that CCS-DIV works with a wide range α and it effectively increases the slope of the "learning" curvature by decreasing α. In contrast, the C-DIV and α-DIV work only for $|\alpha| \leq 1$. However, the convexity of the CCS-DIV will shrink to a small range as α decrease. Thus, we choose the CCS-DIV with $\alpha = -1$ as a more suitable contrast function for devising the corresponding ICA algorithms. Furthermore, the flattest curve is obtainable by CCS-DIV with increasing α similar to E-DIV and C-DIV [13] with $\alpha = 1$.

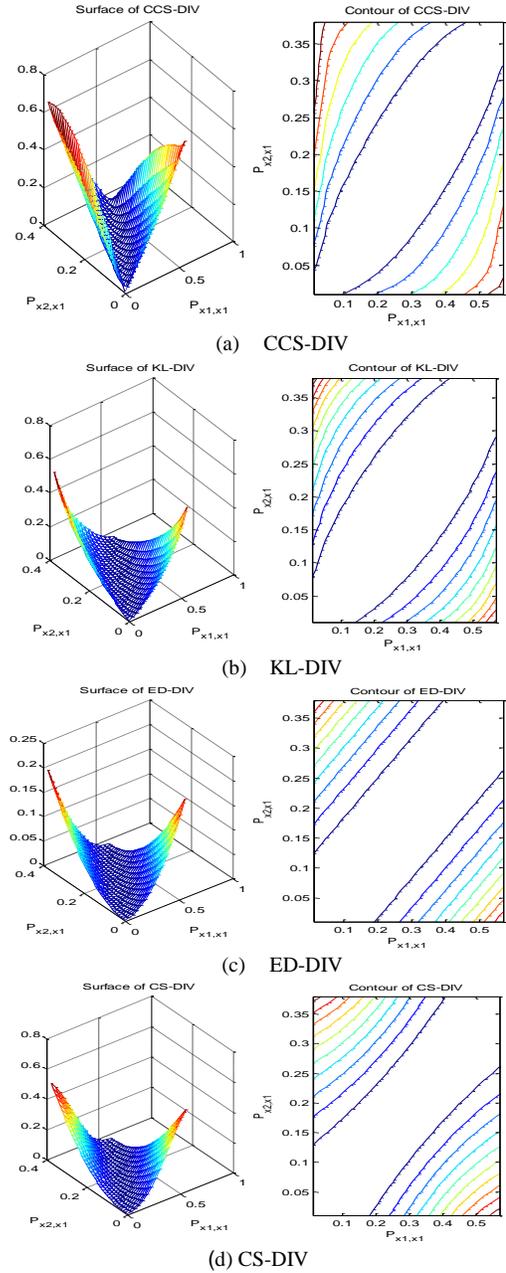

(a) CCS-DIV

(b) KL-DIV

(c) ED-DIV

(d) CS-DIV

Fig.2. The surfaces and Contours of various divergence measures

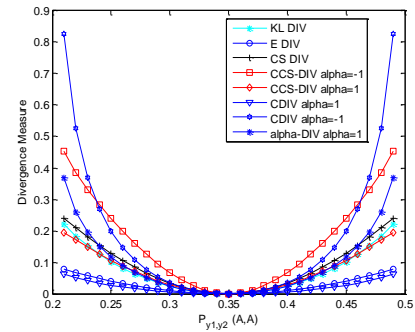

Fig.3. Different divergence measures versus the joint probability $P_{x_1,x_2}(A,A)$

## III. THE CONVEX CAUCHY–SCHWARZ DIVERGENCE INDEPENDENT COMPONENT ANALYSIS (CCS–ICA) PARAMETRIC AND NON-PARAMETRIC ICA ALGORITHMS

### A. Parametric and Non-Parametric ICA algorithms

Without loss of generality, we develop the ICA algorithm by using the CCS-DIV as a contrast function. Let us consider a simple system that is described by the vector-matrix form

$$\mathbf{x} = \mathbf{Hs} + \mathbf{v} \tag{20}$$

where $\mathbf{x} = [x_1, \ldots, x_M]^T$ is a mixture observation vector, $\mathbf{s} = [s_1, \ldots, s_M]^T$ is a source signal vector, $\mathbf{v} = [v_1, \ldots, v_M]^T$ is an additive (Gaussian) noise vector, and $\mathbf{H}$ is an unknown full rank $M \times M$ mixing matrix, where M is the number of source signals. To obtain a good estimate, $\mathbf{y} = \mathbf{Wx}$ of the source signals $\mathbf{s}$, the contrast function CCS-DIV should be minimized with respect to the demixing filter matrix $\mathbf{W}$. Thus, the components of $\mathbf{y}$ become least dependent when this demixing matrix $\mathbf{W}$ becomes a rescaled permutation of $\mathbf{H}^{-1}$. Following the standard ICA procedure, the estimated source $\mathbf{y}$ is typically carried out in two steps: 1) the original data $\mathbf{x}$ should be preprocessed by removing the mean $\{E[\mathbf{x}] = 0\}$ and by a whitening matrix $\{\mathbf{V} = \mathbf{\Lambda}^{-1/2}\mathbf{E}^T\}$, where the matrix $\mathbf{E}$ represents the eigenvectors and $\mathbf{\Lambda}$ the eigenvalues matrices of the autocorrelation, namely, $\{\mathbf{R}_{\mathbf{xx}} = E[\mathbf{xx}^T]\}$. Consequently, the whitened data vector $\{\mathbf{x}_t = \mathbf{Vx}\}$ would have its covariance equal to the identity matrix, i.e., $\{\mathbf{R}_{\mathbf{x}_t\mathbf{x}_t} = \mathbf{I}_K\}$. The demixing matrix can be iteratively computed by, e.g., the gradient descent algorithm [2]:

$$\mathbf{W}(k+1) = \mathbf{W}(k) - \gamma \frac{\partial D_{CCS}(\mathbf{X},\mathbf{W}(k))}{\partial \mathbf{W}(k)} \tag{21}$$

where $k$ represents the iteration index and $\gamma$ is a step size or a learning rate. Therefore, the updated term in the gradient descent is composed of the differentials of the CCS-DIV with respect to each element $w_{ml}$ of the $M \times M$ demixing matrix $\mathbf{W}$. The differentials $\frac{\partial D_{CCS}(\mathbf{X},\mathbf{W}(k))}{\partial w_{ml}(k)}$, $1 \le m, l \le M$ are calculated using a probability model and CCS-DIV measures as in [3], [13] and [14]. However, the update procedure (21) will stop when the absolute increment of the CCS-DIV measure meets a predefined threshold value. During the iterations, we should make the normalization step $\mathbf{w}_m = \mathbf{w}_m / \|\mathbf{w}_m\|$ for each row of $\mathbf{W}$, where $\|.\|$ denotes a norm. Please refer to *Algorithm 1* for more details about the algorithm based on gradient descent.

In setting up the CCS–ICA algorithm based on the proposed CCS-DIV measure, namely, $D_{CCS}(\mathbf{x}_1, \mathbf{x}_2, \alpha)$, usually, the vector $\mathbf{x}_1$ corresponds to the observed data and the vector $\mathbf{x}_2$ corresponds to the estimated or expected data. Here, the CCS–ICA algorithm is detailed as follows. Let the demixed signals $\mathbf{y}_t = \mathbf{Wx}_t$ with its *mth* component denoted as $y_{mt} = \mathbf{w}_m \mathbf{x}_t$. Then, the CCS-DIV as the contrast function, with the built-in convexity parameter $\alpha$, is

$$D_{CCS}(\mathbf{y}_t, y_{mt}, \alpha)$$
$$= \log \frac{\iint f^2(p(\mathbf{y}_t))dy_1 \ldots dy_M \cdot \iint f^2(\prod_1^M p(y_{mt}))dy_1 \ldots dy_M}{[\iint f(p(\mathbf{y}_t)) \cdot f(\prod_1^M p(y_{mt}))\, dy_1 \ldots dy_M]^2} \tag{22}$$

For any convex function, we use the Lebesgue measure to approximate the integral with respect to the joint distribution of $y_t = \{y_1, y_2, \ldots, y_N\}$. The contrast function thus becomes

$$D_{CCS}(\mathbf{y}_t, y_{mt}, \alpha)$$
$$= \log \frac{\sum_1^T f^2(p(\mathbf{Wx}_t)) \cdot \sum_1^T f^2(\prod_1^N (p(w_{mt}\mathbf{x}_t)))}{[\sum_1^T f(p(\mathbf{Wx}_t)) \cdot f(\prod_1^N (p(w_{mt}\mathbf{x}_t)))]^2} \tag{23}$$

The adaptive CCS–ICA algorithms are carried out by using the derivatives of the proposed divergence, i.e., $\left(\partial D_{CCS}(\mathbf{y}_t, y_{mt}, \alpha) / \partial w_{ml}\right)$ as derived in Appendix A. Note that in Appendix A, the derivative of the determinant demixing matrix $(\det(\mathbf{W}))$ with respect to the element $(w_{ml})$ equals the cofactor of entry $(m,l)$ in the calculation of the determinant of $\mathbf{W}$, which we denote as $\left(\frac{\partial \det(\mathbf{W})}{\partial w_{ml}} = W_{ml}\right)$. Also the joint distribution of the output is determined by $p(\mathbf{y}_t) = \frac{p(\mathbf{x}_t)}{|\det(\mathbf{W})|}$.

For simplicity, we can write $D_{CCS}(\mathbf{y}_t, y_{mt}, \alpha)$ as a function of three variables.

$$D_{CCS}(\mathbf{y}_t, y_{mt}, \alpha) = \log \frac{V_1 \cdot V_2}{(V_3)^2} \tag{24}$$

Then,

$$\frac{\partial D_{CCS}(\mathbf{y}_t, y_{mt}, \alpha)}{\partial w_{ml}} = \frac{V_1' V_2 + V_1 V_2' - 2V_1 V_2 V_3'}{V_1 V_2 V_3} \tag{25}$$

where

$$V_1 = \sum_{t=1}^T f^2(\mathbf{y}_t), \quad V_1' = \sum_{t=1}^T 2f(\mathbf{y}_t)f'(\mathbf{y}_t)\mathbf{y}_t'$$

$$V_2 = \sum_{t=1}^T f^2(y_{mt}), \quad V_2' = \sum_{t=1}^T 2f(y_{mt})f'(y_{mt})y_{mt}'$$

$$V_3 = \sum_{t=1}^T f(\mathbf{y}_t)\, f(y_{mt}),$$

$$V_3' = \sum_{t=1}^T f'(\mathbf{y}_t)f(y_{mt})\mathbf{y}_t' + \sum_{t=1}^T f(\mathbf{y}_t)f'(y_{mt})y_{mt}'$$

$$\mathbf{y}_t = p(\mathbf{Wx}_t) \text{ and } y_{mt} = \prod_{m=1}^M p(\mathbf{w}_m \mathbf{x}_t)$$

$$\mathbf{y}_t' = \frac{\partial \mathbf{y}_t}{\partial w_{ml}} = -\frac{p(\mathbf{x}_t)}{|\det(\mathbf{W})|^2} \cdot \frac{\partial \det(\mathbf{W})}{\partial w_{ml}} \cdot \text{sign}(\det(\mathbf{W})),$$

where $\frac{\partial \det(\mathbf{W})}{\partial w_{ml}} = W_{ml}$.

$$y_{mt}' = \frac{\partial y_{mt}}{\partial w_{ml}} = \left[\prod_{j=m}^M p(\mathbf{w}_j \mathbf{x}_t)\right] \frac{\partial p(\mathbf{w}_n \mathbf{x}_t)}{\partial (\mathbf{w}_n \mathbf{x}_t)} \cdot x_l.$$

where $x_l$ denotes the *lth* entry of $\mathbf{x}_t$.

In general, the estimation accuracy of a demixing matrix in the ICA algorithm is limited by the lack of knowledge of the accurate source probability densities. However, non-parametric density estimate is used in [1], [13], [15], by applying the effective Parzen window estimation. One of the attributes of the Parzen window is that it must integrate to one. Thus, it is typical to be a pdf itself, e.g., a Gaussian Parzen window, non-Gaussian or other window functions. Furthermore, it exhibits a distribution shape that is data-driven and is flexibly formed based on its chosen Kernel functions.. Thus, one can estimate the density function $p(y)$ of the process generating the $M$-dimensional sample $\mathbf{y}_1, \mathbf{y}_2 \ldots \mathbf{y}_T$ due to the Parzen Window estimator. For all these reasons, a non-parametric CCS–ICA algorithm is also presented by minimizing the CCS-DIV to generate the demixed signals $\mathbf{y} = [y_1, y_2, \ldots, y_M]^T$. The demixed signals are described by the following univariate and multivariate distributions [3],

$$p(\mathbf{y}_m) = \frac{1}{Th} \sum_{t=1}^{T} \vartheta\left(\frac{y_m - y_{mt}}{h}\right) \quad (26)$$

$$p(\mathbf{y}) = \frac{1}{Th^M} \sum_{t=1}^{T} \varphi\left(\frac{y - y_t}{h}\right) \quad (27)$$

where the univariate Gaussian Kernel is

$$\vartheta(u) = (2\pi)^{-\frac{1}{2}} e^{-\frac{u^2}{2}}$$

and the multivariate Gaussian Kernel is

$$\varphi(\mathbf{u}) = (2\pi)^{-\frac{N}{2}} e^{\frac{-1}{2} u^T u}.$$

The Gaussian kernel(s), used in the non-parametric ICA, are smooth functions. We note that the performance of a learning algorithm based on the non-parametric ICA is better than the performance of a learning algorithm based on the parametric ICA. By substituting (26) and (27) with $\mathbf{y}_t = \mathbf{Wx}_t$ and $y_{mt} = \mathbf{w}_m \mathbf{x}_t$ into (23), the nonparametric CCS-DIV becomes

$$D_{CCS}(\mathbf{y}_t, y_{mt}, \alpha) =$$
$$\log \frac{\sum_{t=1}^{T} f^2(p(\mathbf{Wx}_t)) \cdot \sum_{t=1}^{T} f^2\left(\prod_1^M \frac{1}{Th} \sum_{i=1}^{T} \vartheta\left(\frac{\mathbf{w}_m(\mathbf{x}_t - \mathbf{x}_i)}{h}\right)\right)}{[\sum_{t=1}^{T} f(p(\mathbf{Wx}_t)) \cdot f(\prod_1^M \frac{1}{Th} \sum_{i=1}^{T} \vartheta\left(\frac{\mathbf{w}_m(\mathbf{x}_t - \mathbf{x}_i)}{h}\right))]^2}$$
(28)

However, there are two common methods to minimize this divergence function: one is based on the gradient descent approach and the other is based on an exhaustive search such as the Jacobi method. In this section, we will present the derivation of the proposed algorithm in order to use it in the non-parametric gradient descent ICA algorithm, see *Algorithm 1*. Next, we will discuss how to use the non-parametric ICA algorithm based on the Jacobi optimization method. Thus, the derivative of the divergence $D_{CCS}(\mathbf{y}_t, y_{mt}, \alpha)$ in (28) is given as:

$$\frac{\partial D_{CCS}(\mathbf{y}_t, y_{mt}, \alpha)}{\partial w_{ml}} = \frac{V_1' V_2 + V_1 V_2' - 2V_1 V_2 V_3'}{V_1 V_2 V_3} \quad (29)$$

Where

$$V_1 = \sum_{t=1}^{T} f^2(\mathbf{y}_t), \quad V_1' = \sum_{t=1}^{T} 2f(\mathbf{y}_t)f'(\mathbf{y}_t)\mathbf{y}_t'$$

$$V_2 = \sum_{t=1}^{T} f^2(y_{mt}), \quad V_2' = \sum_{t=1}^{T} 2f(y_{mt})f'(y_{mt})y_{mt}'$$

$$V_3 = \sum_{t=1}^{T} f(\mathbf{y}_t) f(y_{mt}),$$

$$V_3' = \sum_{t=1}^{T} f'(\mathbf{y}_t)f(y_{mt})\mathbf{y}_t' + \sum_{t=1}^{T} f(\mathbf{y}_t)f'(y_{mt})y_{mt}'$$

$$\mathbf{y}_t = p(\mathbf{Wx}_t)$$

$$\mathbf{y}_t' = \frac{\partial \mathbf{y}_t}{\partial w_{ml}} = -\frac{p(\mathbf{x}_t)}{|\det(\mathbf{W})|^2} \cdot \frac{\partial \det(\mathbf{W})}{\partial w_{ml}} \cdot \text{sign}(\det(\mathbf{W})),$$

where $\frac{\partial \det(\mathbf{W})}{\partial w_{ml}} = W_{ml}$; and $\text{sign}(\cdot)$ is the sign function. Thus

$$y_{mt} = \prod_{m=1}^{M} \frac{1}{Th} \sum_{i=1}^{T} \vartheta\left(\frac{y_m - y_{mi}}{h}\right)$$

$$= \prod_{n=1}^{M} \frac{1}{Th} \sum_{i=1}^{T} \vartheta\left(\frac{\mathbf{w}_m(\mathbf{x}_t - \mathbf{x}_i)}{h}\right)$$

$$y_{mt}' = \frac{\partial y_{mt}}{\partial w_{ml}} = -\frac{1}{Th} \sum_{i=1}^{T} \vartheta\left(\frac{\mathbf{w}_m(\mathbf{x}_t - \mathbf{x}_i)}{h}\right) \cdot \left(\frac{\mathbf{w}_m(\mathbf{x}_t - \mathbf{x}_i)}{h}\right)$$

$$\cdot \left(\frac{x_{tl} - x_{il}}{h}\right) \cdot \left[\prod_{j \neq m}^{M} p(\mathbf{w}_j \mathbf{x}_t)\right].$$

where $x_{tl}$ and $x_{il}$ denote the lth entry of $\mathbf{x}_t$.

---

**Algorithm 1:** *ICA Based on the gradient descent*

---

**Input:** $(M \times T)$ *matrix of realizations* $X$, *Initial demixing matrix* $W = I_M$, *Max. number of iterations* $Itr$, *Step Size* $\gamma$ *i.e.* $\gamma = 0.3$, *alpha* $\alpha$ *i.e.* $\alpha = -0.99999$

**Perform Pre-Whitening**
  $\{X = V * X = \Lambda^{\wedge}(-1/2) E^{\wedge}T X\}$,
  **For loop:** *for each I Iteration do*
    **For loop:** *for each* $t = 1, \ldots, T$
      *Evaluate the proposed contrast function and its derivative* $\left(\partial D_{CCS}(Y_t, y_{mt}, \alpha) / \partial w_{ml}\right)$
    **End For**
    *Update de-mixing matrix* $W$
    $$W = W - \gamma \frac{\partial D_{CCS}(X, W)}{\partial W}$$
    *Normalization of* $W$
    *Check Convergence*
    $\|\Delta D_c\| \leq \epsilon$ *i.e.* $\epsilon = 10^{-4}$
  **End For**

**Output:** *Demixing Matrix* $W$, *estimated signals* $y$

---

### B. Scenario of two or three source signals

Generally Speaking, the non-parametric ICA algorithm suffers from insufficient data and high computation in a high dimensional space, especially when estimating the joint distribution. However, in several previous reports in the literature, e.g., [13], [16], the authors suggest applying the pairwise iterative schemes to tackle the high dimensional data problem for non-parametric ICA algorithm(s). However, there are no results indicating how the performance would hold up with the pairwise scheme, especially in terms of computational complexity and in terms of the accuracy of the non-parametric ICA algorithm.

In this work, we present two effective pairwise ICA algorithms: one is based on the gradient descent and the other is based on the Jacobi optimization [16].

Without loss of generality, one can represent the demixing matrix **W** as a series of rotational matrices in terms of unknown angle(s) $\theta_{ij} \in [-\pi/4, \pi/4]$ between each two pair (i, j) of the observed signals. Specifically, define the pairwise rotation matrix

$$W(\theta_{ij}) = \begin{bmatrix} \cos\theta_{ij} & -\sin\theta_{ij} \\ \sin\theta_{ij} & \cos\theta_{ij} \end{bmatrix} \quad (30)$$

The idea is to make each pair of the estimated (marginal) output "independent" as possible (minimize dependency). It was proved and pointed out by Comon in [6] that the mutual independence between the M whitened observed signals might be attained by maximize the independence between each pair of them. In this work, we present two algorithms to solve the high dimensional problem in the non-parametric scheme. First, we adopt the non-parametric algorithm based on the gradient descent into the pairwise iterative scheme of *Algorithm 2*.

Second, we proposed a CCS-ICA algorithm based on Jacobi pairwise scheme in Algorithm 3. This algorithm based on finding the rotation matrix in (28) that attains the minima of CCS-DIV. So, in fact, we set up the range of thetas, such that $\theta_{ij} \in [-\frac{pi}{4}:\theta_g:\frac{pi}{4}]$, where $\theta_g$ is the grid search, for instance $\theta_g = \frac{pi}{64}$. Then for each pair (i, j) of the observation data, we find the demixing matrix $W_2$, which attains the minimum of the CCS-DIV. Please refer to *Algorithm 3* for more details.

---

*Algorithm 2:* **ICA Based on pairwise gradient decent scheme**

**Input:** (M x T) matrix of realization **X**, Initial demixing matrix $W = I_M$, number of iterations Itr, Step Size $\gamma$ i.e. $\gamma = 0.3$, alpha $\alpha$ i.e. $\alpha = -0.99999$
**For** itr $= 1 \ldots$ itrmax
 **Perform Pre-Whitening**
    $\{X = V * X = \Lambda^{((-1)/2)} E^T X\}$,
 **For loop:** for each $i = 1 \ldots M-1$
  **For loop:** for each $j = i+1 \ldots M$
  Initial demixing matrix $W_2 = I_2$
   **While:** while (true)
    Find $W_2$ from due to **Algorithm 1** for each pairs of **X** ;
   **End While**
   Initial rotational matrix
        $R = I_M$,
   Update rotational matrix
        $R([i\ j], [i\ j]) = W_2$
   Update Demixing matrix
        $W = R * W$
  **End For j**
 **End For i**
**End For itr**
**Output:** Demixing matrix $W = W * V$ and demixed sources in $X = W * X$

---

C. *Computational Complexity*

Given T realizations of M observation signals, the

---

*Algorithm 3:* ICA Based on pairwise Jacobi scheme

**Input:** (M x T) matrix of realization X, Initial demixing matrix $W = I_M$, number of iterations Itr, Step Size $\gamma$ i.e. $\gamma = 0.3$, alpha $\alpha$ i.e. $\alpha = -0.99999$
**Perform Pre-Whitening**
    $\{X = V * X = \Lambda^{((-1)/2)} E^T X\}$,
 **While (True)**
  **For loop:** for each $i = 1 \ldots M-1$
   **For loop:** for each $j = i+1 \ldots M$
   **If** $CM([i\ j], [i\ j]) == 0$
        Continue;
   end
   **For loop:** For each $\theta_1 = -\frac{pi}{4}:\frac{pi}{64}:\frac{pi}{4}$
        $W_2 = \begin{bmatrix} \cos\theta_1 & -\sin\theta_1 \\ \sin\theta_1 & \cos\theta_1 \end{bmatrix}$
   Evaluate
        $D_c(X([i\ j],:), W_2 * X([i\ j],:), \alpha)$ For all $t = 1, \ldots, T$.
   **End For**
   Find
        $W_2 = \min_{W_2} D_c(X(i:j,:), W_2 * X, \alpha)$
   Initial rotational matrix
        $R = I_M$,
   Update rotational matrix
        $R([i\ j], [i\ j]) = W_2$
   Update Demixing matrix
        $W = R * W$
   Update Convergence matrix
        $CM([i\ j], [i\ j]) = theta * \frac{180}{pi}$
  **End For**
 **End For**
**End while loop If** $sum(CM) <= 1$
  **Output:** Demixing matrix
   $W = W * V$ and estimated Sources in $X = W * X$

---

computational complexity of the proposed algorithms rely on $T$ and the number of observation signals $M$, and approximately is given by $O\left(\frac{M(M-1)}{2} T^2\right)$. The computational complexity has been a measure of merit for ICA algorithms. With the advent of Graphics Processing Units (GPUs) (see Nvidia.com, e.g.), and more powerful computing platforms, performance accuracy holds more merit. In our comparison among the ICA algorithms, we employ several metrics including computational load time and accuracy. In this work, we employ the adaptive sampling technique that produces improved performance in terms of accuracy and computational load together. The presented technique samples the signal into small time blocks in order to evaluate the integration of the proposed divergence and reduce the computational complexity. Thus, we have introduced sampling factor $T_s$ to evaluate the proposed divergence at each $T_s$ instance. Therefore, the computational complexity of the proposed algorithm is reduced by the square of the sample factor $T_s$ to be less than $O\left(\frac{M(M-1)}{2}\left(\frac{T}{T_s}\right)^2\right)$. Namely, we quantize the specific area of integration of the proposed divergence into equal $\left(\frac{T}{T_s}\right)$ segments to evaluate the proposed divergence.

*Table I:* *The performance of the ICA algorithm based on the proposed divergence and other widely used ICA algorithms in terms of Amari error [2] (multiplied by 100). Each entry averages over the corresponding number of trials. Observation mixtures consists of two source signals that follow the same distribution as denoted in the corresponding example.*

| Source | Samples | Trials | FastICA | JADE | RobustICA | Rapid ICA | IK-DIV | CS-DIV | KL-ICA | ED-DIV | C-DIV | CCS-DIV2 | CCS-DIV3 |
|---|---|---|---|---|---|---|---|---|---|---|---|---|---|
| $s_1, s_1$ | 1000 | 100 | 4.7 | 4.1 | 4.6 | 4.3 | 2.7 | 2.2 | 2.1 | 2.3 | 1.8 | 1.7 | 2.4 |
| $s_2, s_2$ | 1000 | 100 | 6.5 | 4.9 | 6.3 | 6.5 | 3.1 | 2.8 | 2.7 | 2.8 | 2.5 | 2.3 | 1.1 |
| $s_3, s_3$ | 1000 | 100 | 8.2 | 5.6 | 9.3 | 5.8 | 2.2 | 1.8 | 1.8 | 1.9 | 1.9 | 1.5 | 1.5 |
| $s_4, s_4$ | 1000 | 100 | 7.1 | 5.8 | 8.3 | 6.1 | 4.1 | 1.7 | 1.8 | 1.7 | 1.3 | 1.2 | 1.2 |
| $s_1, s_2$ | 1000 | 100 | 3.5 | 3.1 | 3.5 | 3.3 | 2.2 | 1.1 | 1.1 | 1.2 | 1.0 | 0.9 | 0.9 |

*Table II:* *The computational load, in seconds, of the ICA algorithm based on the proposed divergence and other widely used ICA algorithms, each entry averages over the corresponding number of trials. Observation mixtures consists of two source signals that follow the same distribution as denoted in the corresponding example.*

| Source | Samples | Trials | FastICA | JADE | RobustICA | Rapid ICA | IK-DIV | CS-DIV | KL-ICA | ED-DIV | C-DIV | CCS-DIV2 | CCS-DIV3 |
|---|---|---|---|---|---|---|---|---|---|---|---|---|---|
| $s_1, s_1$ | 1000 | 100 | 0.0 | 0.0 | 0.0 | 0.0 | 20.1 | 22.1 | 19.5 | 20.1 | 24.1 | 22.2 | 19.3 |
| $s_2, s_2$ | 1000 | 100 | 0.0 | 0.1 | 0.0 | 0.0 | 20.1 | 21.3 | 19.2 | 20.2 | 23.3 | 19.1 | 21.2 |
| $s_3, s_3$ | 1000 | 100 | 0.0 | 0.0 | 0.0 | 0.0 | 19.1 | 20.7 | 19.1 | 22.1 | 25.1 | 18.1 | 20.2 |
| $s_4, s_4$ | 1000 | 100 | 0.0 | 0.1 | 0.0 | 0.0 | 20.4 | 24.3 | 19 | 23.1 | 24.1 | 19.1 | 19.2 |
| $s_1, s_2$ | 1000 | 100 | 0.0 | 0.0 | 0.0 | 0.0 | 20.2 | 20.1 | 20.1 | 22.1 | 21.4 | 18.1 | 19.2 |

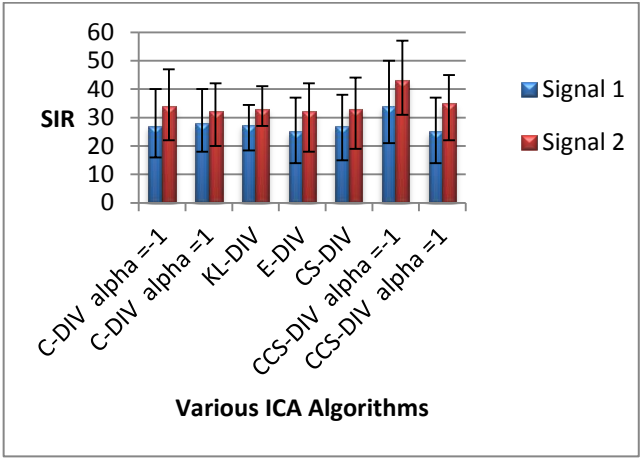

Fig. 4. Comparison of SIRs (dB) of demixed signals by using different ICA algorithms in parametric BSS task.

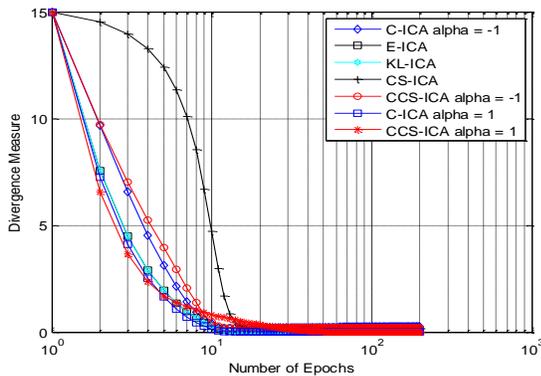

Fig. 5. Comparison of learning curves of C-ICA, E-ICA, KL-ICA, and CCS-ICA with α=1, and α=-1 in a two-source BSS task.

## IV. SIMULATION RESULTS

### A. Performance evaluation of the proposed CCS-ICA algorithms versus the existing ICA-based algorithms

In this section, Monte Carlo Simulations are carried out. It is assumed that the number of sources is equal to the number of observations "sensors". All algorithms have used the same whitening method. The experiments have been carried out using the MATLAB software on an Intel Core i5 CPU 2.4-GHz processor and 4G MB RAM. Each entry corresponds to the average of corresponding trial "independent Monte Carlo" runs in which the mixing matrix is randomly chosen.

First, we compare the performance and convergence speed of the gradient descent ICA algorithms based on the CCS-DIV, CS-DIV, E-DIV, KL-DIV, and C-DIV with $\alpha = 1$ and $\alpha = -1$. In all tasks, the standard gradient descent method is used to devise the parameterized and non-parameterized ICA algorithms based on CCS-DIV with γ=0.7 and γ=0.3 for α=1 and α=-1 cases, respectively, CS-DIV with γ=0.3, E-DIV with γ=0.06, KL-DIV γ=0.17 as in [14], and C-DIV with γ=0.008 and γ=0.1 for α=-1 and α=1 cases, respectively as in [13]. During the comparison, we use bandwidth as a function of sample size, namely, $h = 1.06T^{\frac{-1}{5}}$ [13-15]. To study the parametric scenario for the ICA algorithms, we use mixed signals that consist of two signal sources with a mixing matrix $A = [[0.5\ 0.6]^T [0.3\ 0.4]^T]$, which has a determinant $\det(A) = 0.02$. One of the signal sources has a uniform distribution (sub-Gaussian) and the other has a Laplacian distribution with kurtosis values $-1.2109$ and $3.0839$, respectively. T = 1000 sampled data are taken using a learning rate γ = 0.3 and for 250 iterations. The gradient descent ICA algorithms based on the CCS-DIV, CS-DIV, E-DIV, KL-DIV, and C-DIV with $\alpha = 1$ and $\alpha = -1$, respectively, are implemented to recover the estimated source signals. The initial demixed matrix W is taken as an identity matrix. Fig. 4 shows the demixed signals resulting from the application of the various ICA-based algorithms. Clearly, the parameterized CCS–ICA algorithm outperforms all other ICA algorithms in this scenario with signal to interference ratio (SIR) of 41.9 dB and 32 dB, respectively. Additionally, Fig. 5 shows the "learning curves" of the parameterized CCS–ICA algorithm with α = 1 and α = −1 when compared to the other ICA algorithms, as it graphs the DIV measures versus the iterations (in epochs). As shown in Fig. 5, the speed convergence of the CCS–ICA algorithm is comparable to the C-ICA and KL-ICA algorithms. Furthermore, *Table I* and *II* summarize the performance of the proposed non-parametric ICA algorithms with $\alpha = -1$ against other several algorithms, i.e. CS-DIV, E-DIV, KL-DIV, C-DIV with $\alpha = -1$ and IK-DIV in terms of accuracy and computational complexity, respectively. CCS2 and CCS3 represent *Algorithm 2* and *Algorithm 3*, respectively. We also compare it with other benchmark algorithms such as FastICA [8], RobustICA [7], JADE [11] and RapidICA [42]. For these methods, the default setting parameters are used according to their toolboxes and their publications. In this task, we have examined the aforementioned ICA algorithms to separate mixtures of two sub-Gaussians, two sup-Gaussians, and both sub and sup- Gaussian signals. We use the following distributions: For the sub-Gaussian distribution, we use the uniform distribution

$$p(s_1) = \begin{cases} \frac{1}{2\tau_1} & s_1 \text{ in } (-\tau_1, \tau_1) \\ 0 & \text{Otherwise} \end{cases} \quad (31)$$

and the Rayleigh distribution, we use the following

$$p(s_2) = s_2 \exp\left[-\frac{s_2^2}{2}\right] \quad (32)$$

For the super-Gaussian distribution, we use the Laplacian distribution

$$p(s_3) = \frac{1}{2\tau_2} \exp\left[-\frac{|s_3|}{\tau_2}\right] \quad (33)$$

and log-normal distribution, we use the following

$$p(s_4) = \exp\left[-\frac{(\log s_4)^2}{2}\right] \quad (34)$$

Also, data samples, $T = 1000$, are selected and randomly generated by using $\tau_1 = 3$ and $\tau_2 = 1$. Kurtoses for all aforementioned signals are $-1.2, 2.99, -0.7224,$ and $8.4559$ respectively, and they are evaluated using $\text{Kurt}(s) = E[s^4] / (E[s^2])^2 - 3$.

*Table III*: Kurtosis Values of the different probability density functions that used in the ICA experiments

| Signals' Notation | Kurtosis | Signals' Notation | Kurtosis |
|---|---|---|---|
| $s_1$ | −1.2116 | $s_{12}$ | −0.65419 |
| $s_2$ | 2.9324 | $s_{13}$ | −0.33421 |
| $s_3$ | −1.3995 | $s_{14}$ | −1.6935 |
| $s_4$ | 136.0108 | $s_{15}$ | −0.86239 |
| $s_5$ | 11.6452 | $s_{16}$ | −0.60566 |
| $s_6$ | 4.219 | $s_{17}$ | −0.75488 |
| $s_7$ | −1.2065 | $s_{18}$ | −0.65645 |
| $s_8$ | 3.1965 | $s_{19}$ | −0.81022 |
| $s_9$ | 3.4302 | $s_{20}$ | −0.7692 |
| $s_{10}$ | −1.3049 | $s_{21}$ | −0.27737 |
| $s_{11}$ | −1.6805 | $s_{22}$ | −0.56816 |

One can observe several patterns from *Table I* and *II*. The presented algorithms based on the proposed measure show the best performance in terms of accuracy (in most cases). The proposed algorithm exhibits the comparable behavior in terms of speed with KL and ED. Also, the JADE algorithm performs better than each of FastICA, RobustICA and Rapid ICA in terms of accuracy, but in terms of speed, we find that these later algorithms outperform the JADE algorithm, especially the rapid ICA and Robust ICA.

*Table IV* summarizes the performance of the aforementioned algorithms in a more complex separation process. A different, randomly generated source signals (refer to *Table III*) and mixing matrices are employed. The demixing matrix has been initialized as an identity i.e., $W =$ $I_M$ for all algorithms. As a result, *Table IV* summarizes the performance of each algorithm in terms of the standard error metric (multiplied × 100), see [2]. All results have been averaged over a number of independent Monte Carlo runs. Table IV demonstrates again that the non-parametric ICA based on the proposed divergence provides the best performance in terms of accuracy (in most cases). However, in terms of speed, RapidICA, FastICA, RobustICA and JADE perform better. So, these algorithms could be chosen to initialize for methods of higher performance in order to reduce the overall computational load. Since, the comparison between the ICA algorithms has relied on two criteria, namely, accuracy and computational load, a tradeoff between these two criteria has always been assessed for each targeted application. We also note that with the advent of Graphics Processing Units (GPUs), computational load/speed becomes less of a factor, and the true metric becomes accuracy. *Table* V summarizes the performance of CCS-ICA (see **Algorithm 3**) based on the different values of $T_s$ (1,10,100,1000), and *Table* VI shows their corresponding computational load in seconds. Based on these results, one observes that the best performance of the CCS-ICA, **Algorithm 3**, in terms of accuracy and speed occurs with $T_s = 100$.

**Table IV:** *The performance of the ICA algorithm based on the proposed divergence and other widely used ICA algorithms in terms of Amari error [2] (multiplied by 100). Each entry averages over the corresponding number of trials.*

| Dimensions | Samples | Trials | JADE | FastICA | RapidICA | RobustICA | CS | CDIV | KLDIV | CCS2 | CCS3 |
|---|---|---|---|---|---|---|---|---|---|---|---|
| 2 | 1000 | 512 | 5.6 | 7.3 | 6.1 | 7.2 | 2.5 | 2.2 | 2.3 | 2.1 | 2 |
|   | 2000 | 512 | 5.1 | 5.9 | 5.5 | 6 | 1.9 | 1.7 | 1.9 | 1.8 | 1.8 |
|   | 4000 | 512 | 3.1 | 4.1 | 3.5 | 4.3 | 1.7 | 1.6 | 1.5 | 1.6 | 1.4 |
|   | 8000 | 512 | 2.4 | 2.6 | 2.5 | 2.6 | 1.3 | 1.2 | 1.1 | 1.4 | 1.1 |
| 4 | 1000 | 200 | 8 | 9.7 | 9.1 | 9.8 | 3.0 | 2.4 | 3.1 | 3.1 | 2.5 |
|   | 2000 | 200 | 5.4 | 7.3 | 6.5 | 7.2 | 2.4 | 2.2 | 2.1 | 2.5 | 1.8 |
|   | 4000 | 200 | 4.2 | 4.2 | 4.1 | 4.3 | 1.7 | 1.4 | 1.4 | 1.4 | 1.6 |
|   | 8000 | 200 | 2.1 | 2.7 | 2.5 | 2.7 | 1.4 | 1.2 | 1.3 | 1.4 | 1.2 |
| 8 | 1000 | 75 | 10.5 | 10.3 | 9.6 | 11.2 | 4.6 | 3.6 | 4.2 | 4.4 | 3.2 |
|   | 2000 | 75 | 8.1 | 8.0 | 7.6 | 8.2 | 3.5 | 3.1 | 3.3 | 3.2 | 3 |
|   | 4000 | 75 | 5.7 | 4.1 | 4.4 | 4.9 | 2.5 | 2.3 | 2.7 | 2.6 | 2.8 |
|   | 8000 | 75 | 2.7 | 3.1 | 3.0 | 3.2 | 2.3 | 2.1 | 2 | 2.1 | 1.9 |
| 16 | 1000 | 15 | 8 | 9.7 | 9.1 | 9.8 | 6.7 | 6 | 6.7 | 7.3 | 5.5 |
|   | 2000 | 15 | 5.4 | 7.3 | 6.5 | 7.2 | 6.1 | 5.2 | 6 | 6.9 | 5.1 |
|   | 4000 | 15 | 4.2 | 4.2 | 4.1 | 4.3 | 5.4 | 4.4 | 5.1 | 5.6 | 4.2 |
|   | 8000 | 15 | 2.1 | 2.7 | 2.5 | 2.7 | 3.6 | 2.6 | 3.1 | 3.8 | 2.9 |
| 20 | 1000 | 5 | 22.3 | 21.1 | 20.1 | 26.2 | 14.1 | 9.1 | 10.1 | 13.1 | 8.9 |
|   | 2000 | 5 | 15.7 | 15.6 | 15.2 | 16.2 | 7.7 | 6.7 | 7.3 | 8.3 | 7.2 |
|   | 4000 | 5 | 7.8 | 7.2 | 7.1 | 7.2 | 6.2 | 7.6 | 6.4 | 6.7 | 5.3 |
|   | 8000 | 5 | 4.5 | 4.1 | 3.9 | 4.0 | 2.7 | 2.2 | 2.6 | 4.4 | 2.3 |

### B. Experiments on Speech and Music Signals

Two experiments are documented in this section to evaluate the CCS–ICA in **Algorithm 1**. Both experiments are carried out involving speech and music signals under different conditions. The source signals are two speech signals of different male speakers and a music signal. The first experiment is to separate the three source signals from their mixtures given by $\mathbf{x} = \mathbf{As}$ where the 3 x 3 mixing matrix is

$A =$
$[[0.8 \ 0.3 \ -0.3]^T \ [0.2 \ -0.8 \ 0.7]^T \ [0.3 \ 0.2 \ 0.3]^T \ ]$.

The three speech signals are sampled from the ICA '99 conference BSS test sets at http://sound.media.mit.edu/ica-bench/ [13], [15] with an 8 kHz sampling rate. The non-parameterized CCS–ICA algorithms with $\alpha = 1$ and $\alpha = -1$, (as well as the other existing algorithms), are applied to this task. The resulting waveforms are acquired and the signal to interference ratio (SIR) of each estimated source is calculated. We use the following to calculate the SIR:

Given the source signals $\mathbf{s} = \{s_1, s_2, \dots s_T\}$ and demixed signals $\mathbf{y} = \{y_1, y_2, \dots y_T\}$, the SIR in decibels is calculated by

$$\text{SIR (dB)} = 10 \log \frac{\sum_{t=1}^{T} \|s_t\|^2}{\sum_{t=1}^{T} \|y_t - s_t\|^2}$$
(35)

The summary results are shown in Fig. 6, which also include the SIRs for the other algorithms, namely, JADE, Fast ICA, Robust ICA, KL-ICA and C-ICA with $\alpha = 1$ and $\alpha = -1$. As shown in Fig. 6, the proposed CCS–ICA algorithm achieves significant improvements in terms of SIRs. As exhibited in the previous figures and tables also, the proposed algorithm has consistency and obtains the best performance among the host of listed algorithms. Moreover, a second experiment is conducted to examine the comparative performance in the presence of additive (Gaussian) noise. To that end, we consider the mixing model $\mathbf{x} = \mathbf{As} + \mathbf{v}$ which contains the same source signals with the additive noise and with a different mixing matrix, i.e.
$\mathbf{A} = [[0.8 \ 0.3 \ -0.3]^T \ [0.2 \ -0.8 \ 0.7]^T \ [0.3 \ 0.2 \ 0.3]^T \ ]$

The (Gaussian) noise $\mathbf{v}$ is an $M$ vector with zero mean and $\sigma^2 I$ covariance matrix. In addition, it is independent of the source signals. Fig. 7 shows the SNR of the separated source signals in the noisy BSS model with $SNR = 20\ dB$. Clearly, the proposed **Algorithm 1** has the best performance when compared to others even though its performance decreased in the noisy BSS model. Notably, the SNRs of JADE, Fast ICA and Robust ICA were very low as they rely on the criterion of non-Gaussianity, which is usually less reliable in the Gaussian-noise environment. In contrast, C-ICA, KL-ICA, and the proposed algorithm, which are based on different mutual information measures, achieved reasonable results. We note that one can use the CCS-DIV to recover source signals from the convolutive mixtures in the frequency domain as in [36], [37], and [39].

**Table V:** *The performance of the ICA algorithm based on the proposed divergence in terms of Amari error [2] (multiplied by 100). Each entry averages over the corresponding number of trials.*

| Dimensions M | Samples T | Trials | CCS3 at 0.1T | CCS3 At 0.01T | CCS3 At 0.001T | CCS3 At 1 |
|---|---|---|---|---|---|---|
| 2 | 1000 | 1024 | 4.6 | 2.9 | 2.1 | 2 |
|   | 2000 | 1024 | 3.6 | 2.3 | 1.9 | 1.8 |
|   | 4000 | 1024 | 2.8 | 1.9 | 1.6 | 1.4 |
|   | 8000 | 1024 | 2.2 | 1.6 | 1.1 | 1.2 |
| 4 | 1000 | 250 | 5.8 | 3.8 | 2.4 | 2.5 |
|   | 2000 | 250 | 5 | 2.9 | 2 | 1.8 |
|   | 4000 | 250 | 3.5 | 2.5 | 1.6 | 1.6 |
|   | 8000 | 250 | 2.7 | 2.2 | 1.3 | 1.3 |
| 8 | 1000 | 100 | 5.6 | 3.8 | 2.5 | 3.2 |
|   | 2000 | 100 | 3.7 | 3.1 | 2.2 | 3 |
|   | 4000 | 100 | 3.1 | 2.6 | 2.2 | 2.8 |
|   | 8000 | 100 | 3.0 | 2.2 | 1.9 | 1.9 |
| 16 | 1000 | 25 | 20.5 | 15.8 | 8.6 | 5.5 |
|   | 2000 | 25 | 12.6 | 10.1 | 7 | 5.1 |
|   | 4000 | 25 | 8.6 | 8 | 4.5 | 4.2 |
|   | 8000 | 25 | 5.8 | 3.9 | 1.9 | 2.9 |
| 20 | 1000 | 10 | 27.7 | 15.1 | 13.7 | 8.9 |
|   | 2000 | 10 | 22.8 | 11.3 | 12 | 7.2 |
|   | 4000 | 10 | 15.6 | 9 | 7.2 | 5.3 |
|   | 8000 | 10 | 9.8 | 6.3 | 3 | 2.3 |

**Table VI:** *The computational load, in seconds, of the ICA algorithm based on the proposed divergence and other widely used ICA algorithms, each entry averages over the corresponding number of trials.*

| Dimensions M | Samples T | Trials | CCS3 at 0.1T | CCS3 At 0.01T | CCS3 At 0.001T | CCS3 At 1 |
|---|---|---|---|---|---|---|
| 2 | 1000 | 1024 | 0.4 | 2.8 | 29.8 | 28 |
|   | 2000 | 1024 | 0.5 | 4.8 | 44.8 | 96.4 |
|   | 4000 | 1024 | 0.8 | 8 | 77.9 | 342.9 |
|   | 8000 | 1024 | 1.5 | 10.6 | 137 | 1073 |
| 4 | 1000 | 250 | 1.8 | 24 | 218.1 | 237.9 |
|   | 2000 | 250 | 4.3 | 39 | 344.8 | 630.3 |
|   | 4000 | 250 | 5.9 | 47.9 | 593.4 | 2348.6 |
|   | 8000 | 250 | 10.2 | 83.6 | 1105 | 7737.1 |
| 8 | 1000 | 100 | 19.3 | 128.7 | 1053 | 1174 |
|   | 2000 | 100 | 31.5 | 201.7 | 1743 | 3347 |
|   | 4000 | 100 | 46.5 | 266.4 | 3109 | 11705 |
|   | 8000 | 100 | 74.2 | 241.8 | 5534 | 42115 |
| 16 | 1000 | 25 | 170.6 | 909.5 | 6282 | 4376.2 |
|   | 2000 | 25 | 242.3 | 1171 | 9320 | 17918.3 |
|   | 4000 | 25 | 305.5 | 1403 | 14717 | 58894.6 |
|   | 8000 | 25 | 329.9 | 2297 | 25658 | 10483.4 |
| 20 | 1000 | 10 | 339 | 1195.7 | 9605 | 11355.2 |
|   | 2000 | 10 | 427.4 | 1724.2 | 14708 | 27504.8 |
|   | 4000 | 10 | 607.6 | 2398.3 | 23634 | 52536.6 |
|   | 8000 | 10 | 900 | 3754.5 | 42538 | 97312.1 |

## V. CONCLUSION

A novel divergence measure is presented based on integrating convex functions *into* the Cauchy–Schwarz inequality. This divergence measure is used as a contrast function to develop new ICA algorithms to solve the Blind Source Separation (BSS) problem. The CCS-DIV derived algorithms can be controlled to attain the steepest descent towards the minimum value. Also, a pairwise iterative scheme is employed to address the high dimensional problem in BSS. Two schemes of pairwise non-parametric ICA algorithms are developed based on the proposed divergence. Several examples and experiments are carried out to show the improved performance of the proposed divergence. Furthermore, this paper compares the metric performance with a host of leading ICA algorithms. We have developed also nonparametric CCS–ICA approaches to demixing where the source signals are estimated by the Parzen Window density. The convergence speed of the parameterized CCS–ICA procedure is evaluated and compared to other algorithms. The proposed CCS–ICA algorithms attained the highest SIR in separation of speech and music signals relative to other leading ICA-based algorithms.

## APPENDIX A

### CONVEX CAUCHY–SCHWARZ DIVERGENCE AND ITS DERIVATIVE

Assume the demixed signals $\mathbf{y_t} = \mathbf{W}\mathbf{x_t}$ where the $mth$ component is $y_{mt} = \mathbf{w}_m\mathbf{x_t}$. Now, express the CCS-DIV as a contrast function with a convexity parameter $\alpha$ as follows:

$$D_{CCS}(\mathbf{y_t}, y_{mt}, \alpha) = \log \frac{\iint f^2(p(\mathbf{y_t}))dy_1 \dots dy_M \cdot \iint f^2(\prod_1^M p(y_{mt}))dy_1 \dots dy_M}{[\iint f(p(\mathbf{y_t})) \cdot f(\prod_1^M p(y_{mt}))\,dy_1 \dots dy_M]^2}$$

By using the Lebesgue measure to approximate the integral with respect to the joint distribution of $\mathbf{y_t} = \{y_1, y_2, \dots, y_M\}$, the contrast function becomes

$$D_{CCS}(\mathbf{y_t}, y_{mt}, \alpha) = \log \frac{\sum_1^T f^2(p(\mathbf{W}\mathbf{x_t})) \cdot \sum_1^T f^2(\prod_1^M (p(\mathbf{w_m}\mathbf{x_t})))}{[\sum_1^T f(p(\mathbf{W}\mathbf{x_t})) \cdot f(\prod_1^M (p(\mathbf{w_m}\mathbf{x_t})))]^2}$$

For simplicity, let us assume

$$V_1 = \sum_{t=1}^T f^2(\mathbf{y_t}),\quad V_1' = \sum_{t=1}^T 2f(\mathbf{y_t})f'(\mathbf{y_t})y_t'$$

$$V_2 = \sum_{t=1}^T f^2(y_{mt}),\quad V_2' = \sum_{t=1}^T 2f(y_{mt})f'(y_{mt})y_{mt}'$$

$$V_3 = \sum_{t=1}^T f(\mathbf{y_t})\,f(y_{mt}),$$

$$V_3' = \sum_{t=1}^T f'(\mathbf{y_t})f(y_{mt})y_t' + \sum_{t=1}^T f(\mathbf{y_t})f'(y_{mt})y_{mt}'$$

and the convex function is

$$f(t) = \frac{4}{1-\alpha^2}\left[\frac{1-\alpha}{2} + \frac{1+\alpha}{2}t - t^{\frac{1+\alpha}{2}}\right]$$

$$f'(t) = \frac{2}{1-\alpha}\left[1 - t^{\alpha-1/2}\right]$$

then,

$$\mathbf{y_t} = p(\mathbf{W}\mathbf{x_t})\text{ and } y_{mt} = \prod_{m=1}^M p(\mathbf{w_m}\mathbf{x_t})$$

$$\mathbf{y_t}' = \frac{\partial \mathbf{y_t}}{\partial w_{ml}} = -\frac{p(\mathbf{x_t})}{|\det(\mathbf{W})|^2} \cdot \frac{\partial \det(\mathbf{W})}{\partial w_{ml}} \cdot \text{sign}(\det(\mathbf{W})),$$

where $\frac{\partial \det(\mathbf{W})}{\partial w_{ml}} = W_{ml}$;

$$y_{mt}' = \frac{\partial y_{mt}}{\partial w_{ml}} = \left[\prod_{j\neq m}^M p(\mathbf{w_j}\mathbf{x_t})\right]\frac{\partial p(\mathbf{w_m}\mathbf{x_t})}{\partial (\mathbf{w_m}\mathbf{x_t})} \cdot x_l.$$

where $x_l$ denotes the $l^{th}$ entry of $\mathbf{x}_t$.

Thus, we re-write the CCS-DIV as

$$D_{CCS}(\mathbf{y_t}, y_{mt}, \alpha) = \log \frac{V_1 \cdot V_2}{[V_3]^2}$$

and its derivative becomes

$$\frac{\partial D_{CCS}(\mathbf{y_t}, y_{mt}, \alpha)}{\partial w_{ml}} = \frac{V_3^2}{V_1 \cdot V_2} \cdot \frac{V_1'V_2V_3 + V_1V_2'V_3 - 2V_1V_2V_3V_3'}{V_3^4}$$

$$\frac{\partial D_{CCS}(\mathbf{y_t}, y_{mt}, \alpha)}{\partial w_{ml}} = \frac{V_1'V_2 + V_1V_2' - 2V_1V_2V_3'}{V_1V_2V_3}$$

## ACKNOWLEDGMENT

We would like to thank the anonymous reviewers for their helpful comments and evaluation of this paper.

## REFERENCES

[1] P. Comon, C. Jutten (eds.), "Handbook of Blind Source Separation Independent Component Analysis and Applications." (Academic Press, Oxford, 2010).
[2] A. Cichocki, S.-I. Amari, Adaptive Blind Signal and Image Processing: Learning Algorithms and Applications, John Wiley & Sons, Inc., 2002.
[3] A. Cichocki, R. Zdunek, S.-I. Amari, Nonnegative matrix and tensor factorizations: applications to exploratory multi-way analysis and Blind Source Separation, John Wiley & Sons, Inc., 2009.
[4] S. Boyd and L. Vandenberghe Convex Optimization, 2004: Cambridge Univ. Press.
[5] C. E. Shannon "A mathematical theory of communication," Bell Syst. Tech. J., vol. 27, pp. 379 –423, 1948.
[6] P. COMON, ``Independent Component Analysis, a new concept," Signal Processing, Elsevier, 36(3):287--314, April 1994PDF Special issue on Higher-Order Statistics.

[7] V. Zarzoso and P. Comon, "Robust Independent Component Analysis by Iterative Maximization of the Kurtosis Contrast with Algebraic Optimal Step Size," IEEE Transactions on Neural Networks, vol. 21, no. 2, pp. 248–261, 2010.

[8] Hyvarinen. A, "Fast and robust fixed-point algorithm for independent component analysis". IEEE Transactions on Neural Network, vol. 10, no. 3, pp. 626–634, May 1999.

[9] Hyvarinen. A. E.Oja, "A fast fixed-point algorithm for independent component analysis," Neural Computation, vol. 9, no. 7, pp. 1483–1492, 1997.

[10] F. Cardoso. On the performance of orthogonal source separation algorithms. In Proc. EUSIPCO, pages 776–779, 1994a.

[11] Jean-François Cardoso, "High-order contrasts for independent component analysis," Neural Computation, vol. 11, no 1, pp. 157–192, Jan. 1999.

[12] K. Waheed and F. Salem, "Blind information-theoretic multiuser detection algorithms for DS-CDMA and WCDMA downlink systems," IEEE Trans. Neural Netw., vol. 16, no. 4, pp. 937–948, Jul. 2005.

[13] Jen-Tzung Chien, Hsin-Lung Hsieh, "Convex Divergence ICA for Blind Source Separation," Audio, Speech, and Language Processing, IEEE Transactions on, On page(s): 302–313 Volume: 20, Issue: 1, Jan. 2012

[14] D. Xu , J. C. Principe , J. Fisher III and H.-C. Wu "A novel measure for independent component analysis (ICA)," Proc. Int. Conf. Acoust., Speech, Signal Process., vol. 2, pp. 1161–1164, 1998

[15] R. Boscolo , H. Pan and V. P. Roychowdhury "Independent component analysis based on nonparametric density estimation," IEEE Trans. Neural Netw., vol. 15, no. 1, pp. 55–65, 2004.

[16] Y. Chen "Blind separation using convex function," IEEE Trans. Signal Process., vol. 53, no. 6, pp. 2027–2035, 2005.

[17] J.-T. Chien and B.-C. Chen "A new independent component analysis for speech recognition and separation," IEEE Trans. Audio, Speech Lang. Process., vol. 14, no. 4, pp. 1245–1254, 2006

[18] Y. Matsuyama, N. Katsumata, Y. Suzuki and S. Imahara "The $\alpha$-ICA algorithm," Proc. Int. Workshop Ind. Compon. Anal. Blind Signal Separat., pp. 297 –302, 2000

[19] A. Cichocki, R. Zdunek, and S. Amari, " Csiszár's Divergences for Non-Negative Matrix Factorization: Family of New Algorithms," 6th International Conference on Independent Component Analysis and Blind Signal Separation, Charleston SC, USA, March 5–8, 2006 Springer LNCS 3889, pp. 32–39.

[20] H.-L. Hsieh and J.-T. Chien "A new nonnegative matrix factorization for independent component analysis," Proc. Int. Conf. Acoust., Speech, Signal Process., pp. 2026–2029, 2010

[21] E. Moulines, J.-F. Cardoso, and E. Gassiat, "Maximum likelihood for blind separation and deconvolution of noisy signals uses mixture models," in Proc. Int. Conf. Acoust. Speech Signal Process. Apr. 1997, vol. 5, pp. 3617–3620.

[22] D. T. Pham and P. Garat, "Blind separation of mixture of independent sources through a quasi-maximum likelihood approach," IEEE Trans. Signal Process. vol. 45, no. 7, pp. 1712–1725, Jul. 1997.

[23] B. A. Pearlmutter and L. C. Parra, "Maximum likelihood blind source separation: A context-sensitive generalization of ICA," Adv. Neural Inf. Process. Syst., pp. 613–619, Dec. 1996.

[24] Fujisawa, H. and Eguchi, S. (2008). Robust parameter estimation with a small bias against heavy contamination. *J. Multivariate Anal.* **99** 2053–2081.

[25] S.Eguchi, Y.Kano, "robust maximum likelihood estimation," In institute of Statistical Mathematics, Tokyo, (2001).

[26] J. Zhang "Divergence function, duality, and convex analysis," Neural Comput., vol. 16, pp. 159 –195, 2004.

[27] J. Lin "Divergence measures based on the Shannon entropy," IEEE Trans. Inf. Theory, vol. 37, no. 1, pp. 145 –151, 1991.

[28] S.C.Douglas, X.Sun, "Convolutive blind separation of speech mixtures using the natural gradient," Speech commun, vol. 39. pp. 65–78, (2002).

[29] Yoshioka, Takuya Nakatani, Tomohiro Miyoshi, Masato Okuno, Hiroshi G. "Blind Separation and Dereverberation of Speech Mixtures by Joint Optimization," IEEE Transactions on Audio Speech and Language Processing, Volume. 19, Issue. 1, pp. 69, 2011.

[30] A. Cichocki, R. Zdunek, S. Amari, G. Hori and K. Umeno, "Blind Signal Separation Method and System Using Modular and Hierarchical-Multilayer Processing for Blind Multidimensional Decomposition, Identification, Separation or Extraction," Patent pending, No. 2006-124167, RIKEN, Japan, March 2006.

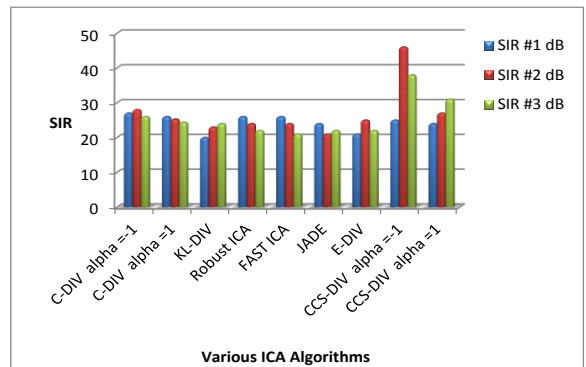

Fig. 6. Comparison of SIRs (dB) of demixed two speeches and music signals by using different ICA algorithms in instantaneous BSS task.

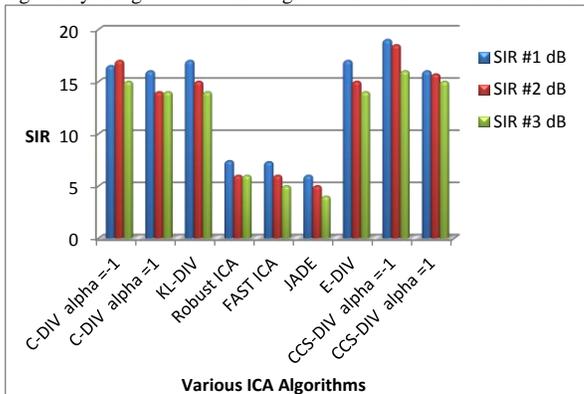

Fig. 7. Comparison of SIRs (dB) of demixed two speeches and music signals by using different ICA algorithms in instantaneous BSS task with additive Gaussian noise.

[31] K. E. Hild, II, D. Erdogmus, and J. C. Principe, "An analysis of entropy estimators for blind source separation," Signal Process., vol. 86, no. 1, pp. 182–194, Jan. 2006.

[32] Ali, S.M., Silvey, S.D., 1966. A general class of coefficient of divergence of one distribution from another. J. Roy. Statist. Soc. 28 (1), 131–142.

[33] M. Minami and S. Eguchi, "Robust Blind Source Separation by beta-Divergence," Neural Computation Vol.14, pp. 859-1886, 2002.

[34] M. N. H. Mollah, M. Minami and S. Eguchi, "Exploring Latent Structure of Mixture ICA Models by the Minimum β-Divergence Method," Neural Computation, Vol. 18(1), pp. 166-190, 2006.

[35] Low, S.Y. Nordholm, S. Togneri, R. "Convolutive Blind Signal Separation With Post-Processing," IEEE Transactions on Speech and Audio Processing, Volume 12, Issue 5, pp. 539, 2004.

[36] Takeda, R., Nakadai, K., Takahashi, T., Komatani, K., Ogata, T., Okuno, H.G., "Step-size parameter adaptation of multi-channel semi-blind ICA with piecewise linear model for barge-in-able robot audition," Intelligent Robots and Systems, 2009. IROS 2009. IEEE/RSJ International Conference on, On page(s): 2277–228.2

[37] Takeda, R., Nakadai, K., Takahashi, T., Komatani, K., Ogata, T., Okuno, H.G., "Upper-limit evaluation of robot audition based on ICA-BSS in multi-source, barge-in and highly reverberant conditions," Robotics and Automation (ICRA), 2010 IEEE International Conference on, On page(s): 4366–4371.

[38] Takahashi, Yu., Saruwatari, H., Shikano, K., "Real-time implementation of blind spatial subtraction array for hands-free robot spoken dialogue system," Intelligent Robots and Systems, 2008. IROS 2008. IEEE/RSJ International Conference on, On page(s): 1687–1692.

[39] Jan, T., Wenwu Wang, DeLiang Wang, "A multistage approach for blind separation of convolutive speech mixtures," Acoustics, Speech and Signal Processing, 2009. ICASSP 2009. IEEE International Conference on, On page(s): 1713–1716.

[40] Yueyue Na, Jian Yu, "Kernel and spectral methods for solving the permutation problem in frequency domain BSS," Neural Networks (IJCNN), The 2012 International Joint Conference on, On page(s): 1 – 8

[41] A. Chen, "Fast kernel density independent component analysis," in Proc. 6th Int. Conf. ICA BSS, 2006, vol. 3889, pp. 24–31.

[42] R. Yokote, and Y. Matsuyama, "Rapid Algorithm for Independent Component Analysis" Journal of Signal and Information Processing, 2012, 3, 275-285.